\documentclass[twocolumn,preprintnumbers,aps,pra,superscriptaddress,amsmath,amssymb]{revtex4-1}

\usepackage{graphicx}
\usepackage{dcolumn}
\usepackage{bm}
\usepackage{bbm}
\usepackage{hyperref}
\usepackage{color}
\usepackage{multirow}
\usepackage{physics}
\DeclareMathOperator{\es}{env}
\DeclareMathOperator{\mw}{MW}
\DeclareMathOperator{\nv}{NV}

\begin{document}

\title{Environmentally mediated coherent control of a spin qubit in diamond}
	
\author{Scott E. Lillie}
\affiliation{Centre for Quantum Computation and Communication Technology, School of Physics, The University of Melbourne, VIC 3010, Australia}

\author{David A. Broadway}
\affiliation{Centre for Quantum Computation and Communication Technology, School of Physics, The University of Melbourne, VIC 3010, Australia}

\author{James D. A. Wood}
\altaffiliation{Present address: Department of Physics, University of Basel, Switzerland}
\affiliation{Centre for Quantum Computation and Communication Technology, School of Physics, The University of Melbourne, VIC 3010, Australia}

\author{David A. Simpson}
\affiliation{School of Physics, The University of Melbourne, VIC 3010, Australia}

\author{Alastair Stacey}
\affiliation{Centre for Quantum Computation and Communication Technology, School of Physics, The University of Melbourne, VIC 3010, Australia}
\affiliation{Melbourne Centre for Nanofabrication, 151 Wellington Road, Clayton, VIC 3168, Australia}

\author{Jean-Philippe Tetienne} 
\email{Corresponding author: jtetienne@unimelb.edu.au}
\affiliation{Centre for Quantum Computation and Communication Technology, School of Physics, The University of Melbourne, VIC 3010, Australia}
\affiliation{School of Physics, The University of Melbourne, VIC 3010, Australia}

\author{Lloyd C. L. Hollenberg}
\affiliation{Centre for Quantum Computation and Communication Technology, School of Physics, The University of Melbourne, VIC 3010, Australia}
\affiliation{School of Physics, The University of Melbourne, VIC 3010, Australia}

\date{\today}
	
\begin{abstract}
	
The coherent control of spin qubits forms the basis of many applications in quantum information processing and nanoscale sensing, imaging and spectroscopy. Such control is conventionally achieved by direct driving of the qubit transition with a resonant global field, typically at microwave frequencies. Here we introduce an approach that relies on the resonant driving of nearby environment spins, whose localised magnetic field in turn drives the qubit when the environmental spin Rabi frequency matches the qubit resonance. This concept of environmentally mediated resonance (EMR) is explored experimentally using a qubit based on a single nitrogen-vacancy (NV) centre in diamond, with nearby electronic spins serving as the environmental mediators. We demonstrate EMR driven coherent control of the NV spin-state, including the observation of Rabi oscillations, free induction decay, and spin-echo. This technique also provides a way to probe the nanoscale environment of spin qubits, which we illustrate by acquisition of electron spin resonance spectra of single NV centres in various settings.

\end{abstract}

\maketitle

The coherent control of spin-state qubits is fundamental to endeavours in both quantum computing and nanoscale sensing. In quantum computing, the ability to coherently control the spin-state of a target qubit within an array is essential to quantum information processing, and to harnessing the enhanced computing power of quantum algorithms \cite{Kane1998,Ladd2010,Pla2012,Hill2015}. In quantum sensing, the coherent control of a qubit spin-state is required to selectively decouple the qubit from its magnetic environment, enhancing sensitivity to some target signal window \cite{Budker2007,Cole2009,Hall2009,Rondin2014,Bienfait2015,Degen2016}. This has led to a significant decrease in sensing volumes as compared to conventional magnetic resonance experiments \cite{Staudacher2013c,Mamin2013}, achieving detection at the single-electron-spin level \cite{Grinolds2013,Shi2015}, and holds promise towards atomic-resolution imaging of single biomolecules \cite{Ajoy2015,Kost2015,Laraoui2015,Lazariev2015,Perunicic2016}.

Coherent control of qubit spin-states is typically achieved directly, by application of a global driving field resonant with a qubit transition, whilst the fluctuating states of spins present in the local qubit environment decohere the qubit state. The unwanted decoherence caused by these environmental spins typically limits the ability to perform complex algorithms for quantum computing, or equivalently, limits the performance of the qubit for sensing purposes. Here we present a technique by which these typically chaotic environmental spins are appropriated as localised agents of control, allowing the coherent manipulation of a proximal qubit state. Precisely, control of the qubit is achieved by matching the Rabi frequency of directly driven environmental spins with the qubit spin-transition frequency. This environmentally mediated resonance (EMR) condition therefore classifies as a Hartmann-Hahn-like double resonance \cite{Hartmann1962,Belthangady2013,London2013,Loretz2013}.

To demonstrate this concept experimentally, we use a qubit based on a single nitrogen-vacancy (NV) defect centre in diamond, which can be optically initialised and read out under ambient conditions \cite{Doherty2013}, and enlist an ensemble of nearby electron spins as environmental mediators. The resonance landscape of EMR is explored by varying driving frequency, driving strength, and external field strength, and is found to be in good agreement with a simple semi-classical model. Coherent control of the NV spin-state is illustrated by performing EMR driven analogues of Rabi, free induction decay, and spin-echo experiments. Finally, applications to nanoscale spectroscopy are demonstrated, including the acquisition of a substitutional nitrogen spectrum by EMR driving.

\begin{figure}[t]
	\begin{center}
		\includegraphics[width=0.5\textwidth]{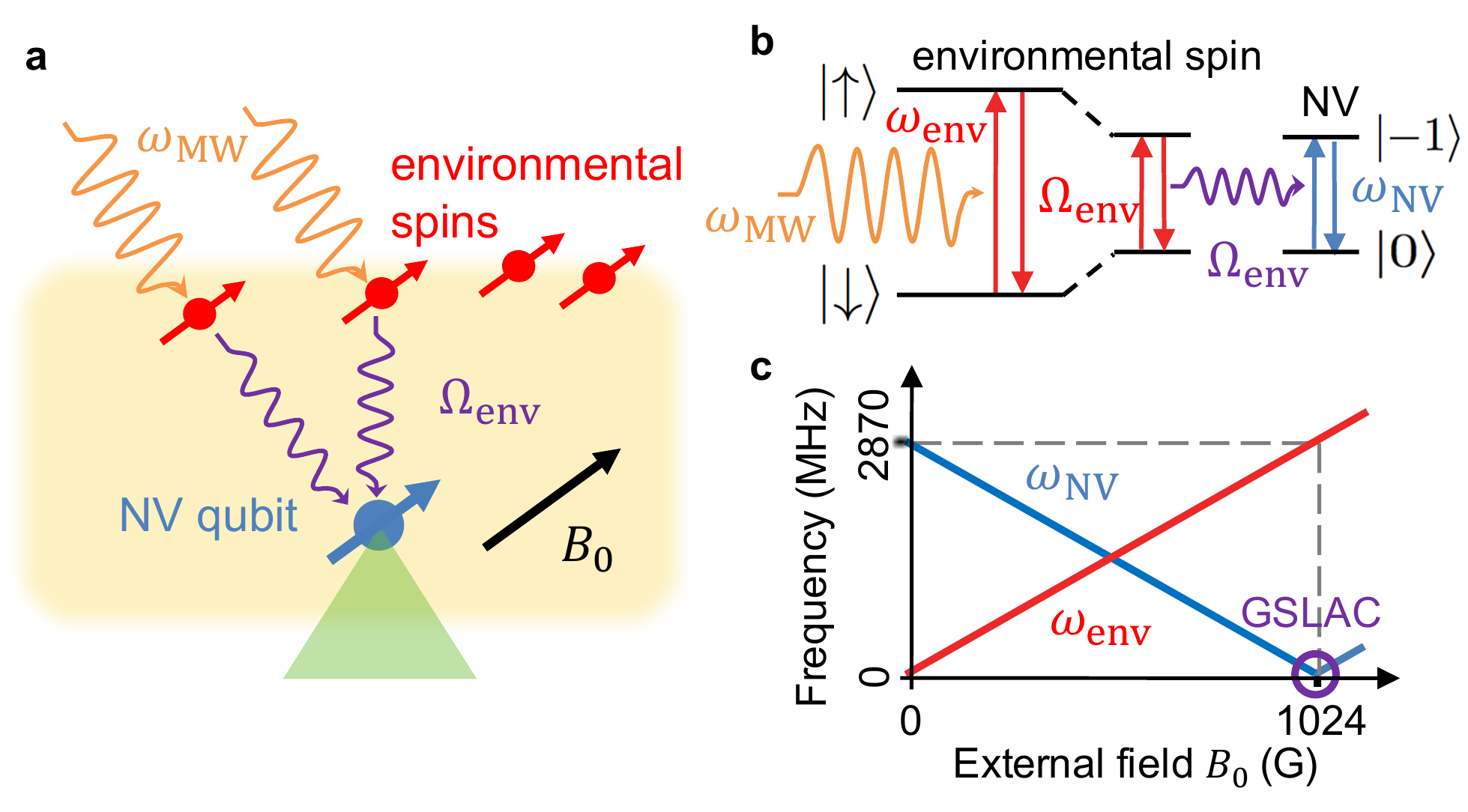}
		\caption{\textbf{a}, Environmentally mediated resonance schematic. Near surface environmental spins (red) are driven directly by a global microwave field (orange) at frequency $\omega_{\text{MW}}\approx\omega_{\es}$, giving rise to a localised oscillating magnetic field (purple) at the environmental spin Rabi frequency, $\Omega_{\es}$. This field in turn drives a proximal NV qubit (blue) when the environmental spin Rabi frequency matches the qubit transition frequency, $\Omega_{\es}\approx\omega_{\text{NV}}$. An external magnetic field, $B_0$, is aligned with the NV quantisation axis, defining the quantisation axis of the environmental spins, and the NV spin-state is read optically (green). \textbf{b}, EMR matching condition requires a directly driven environmental spin Rabi frequency, or equivalently, dressed transition frequency, $\Omega_{\es}$, to be brought into resonance with the qubit transition frequency, $\omega_{\nv}$. \textbf{c}, EMR matching condition is conveniently achieved at an external field strength $B_0\approx1024$ G, giving an NV $\ket{0}\leftrightarrow\ket{-1}$ spin-state transition frequency $\omega_{\text{NV}}<10$ MHz, close to the ground-state level anti-crossing (GSLAC), and a corresponding $\ket{\downarrow}\leftrightarrow\ket{\uparrow}$ free-electron environmental spin transition frequency of $\omega_{\es}\approx2870$ MHz.}
		\label{Fig1}
	\end{center}
\end{figure}

 \begin{figure}[t]
 	\begin{center}
 		\includegraphics[width=0.5\textwidth]{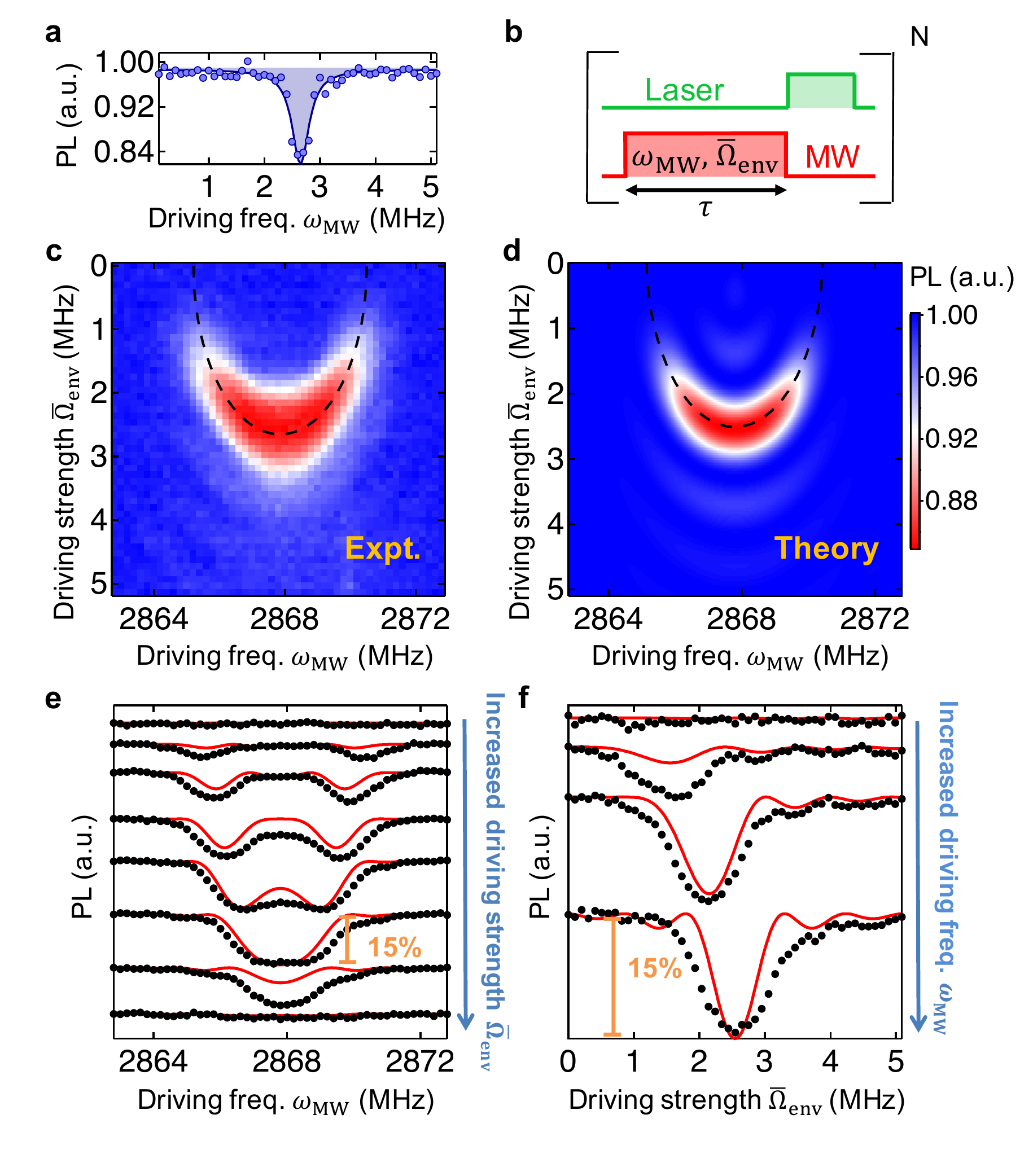}
 		\caption{\textbf{a}, Directly driven magnetic resonance spectrum of a single near surface NV as prepared for the subsequent measurements, at an external field of $B_0\approx1023$ G, giving a single degenerate transition frequency at the $^{15}$NV hyperfine crossing, $\omega_{\text{NV}}=2.65$ MHz. \textbf{b}, Pulse sequence used for these experiments. A microwave pulse of frequency $\omega_{\text{MW}}$, driving strength $\overline{\Omega}_{\es}=\gamma_e B_1/2$, and duration, $\tau$ drives the environmental spin bath, and a single $532$ nm laser pulse is used for optical readout and re-pumping of the NV spin-state. \textbf{c}, Experimental EMR PL map as a function of $\omega_{\text{MW}}$ and $\overline{\Omega}_{\es}$ as measured for the near surface NV presented in Fig. \ref{Fig2}a. Dashed line shows the EMR matching condition, Eq. (\ref{Eqn2}). \textbf{d}, Theoretical EMR PL map as given by Eq. (\ref{Eqn1}), with probability scaled to match the directly driven PL contrast in Fig. \ref{Fig2}a. \textbf{e}, Driving frequency and \textbf{f}, driving strength line cuts of the above EMR PL maps with points taken from the experimental map, and curves from the theory map. Driving strength increases from $1$ MHz to $3.5$ MHz from top to bottom for the frequency line cuts, and driving frequency increases from $2864$ MHz to $2868$ MHz from top to bottom for the driving strength line cuts.} 
 		\label{Fig2}
 	\end{center}
 \end{figure}
 
\paragraph*{Principle -}
 
The resonance landscape of EMR driving can be understood by a simple semi-classical model, formulated by double-application of the Rabi formula, first applied to the environmental spins under direct driving, modelled as a single macrospin for simplicity, and second to the qubit, here the NV centre, as driven by the effective field arising from the environmental spin Rabi oscillations [Fig. \ref{Fig1}a]. Assuming the NV spin is initialised in the $\ket{0}$ state, this model gives the probability of measuring the $\ket{-1}$ state after a driving time, $\tau$, as \cite{SI}
\begin{equation}
P_{\ket{-1}}(\tau) = \left(\dfrac{\overline{\Omega}_{\text{NV}}}{\Omega_{\text{NV}}}\right)^2 \dfrac{1 - \cos(\Omega_{\text{NV}} \tau)}{2} \label{Eqn1}
\end{equation}

\noindent where $\Omega_{\nv} = \sqrt{\overline{\Omega}_{\text{NV}}^2 + (\Omega_{\es} - \omega_{\text{NV}})^2}$ is the Rabi frequency of the NV, with an effective driving strength $\overline{\Omega}_{\text{NV}}=\gamma_e\,\alpha(\{\vec{r}_i\})\, (\overline{\Omega}_{\es}/\Omega_{\es})^2/\sqrt{2}$. The latter is dependent on the relative amplitude of the environmental spin Rabi oscillations, $(\overline{\Omega}_{\es}/ \Omega_{\es})^2$, and the net magnetic field projection perpendicular to the NV-axis due to the ensemble of environmental spins, $\alpha(\{\vec{r}_i\})$, at positions $\{\vec{r}_i\}$ relative to the NV \cite{SI}. The environmental spin Rabi frequency is given by $\Omega_{\es}= \sqrt{\overline{\Omega}_{\es}^2 + (\omega_{\text{MW}} - \omega_{\es})^2}$, with driving strength $\overline{\Omega}_{\es}=\gamma_eB_{1}/2$, where $\gamma_e$ is the electron gyromagnetic ratio and $B_1$ is the direct driving magnetic field amplitude. The NV $\ket{0}\leftrightarrow\ket{-1}$ spin transition frequency is denoted as $\omega_{\text{NV}}$, that of the spin-1/2 environmental spins as $\omega_{\es}$, and the driving microwave field frequency is $\omega_{\text{MW}}$. 

Equation (\ref{Eqn1}) suggests an EMR matching condition when the environmental spin Rabi frequency, or equivalently the dressed state transition frequency, is brought into resonance with the NV transition frequency, $\omega_{\nv}=\Omega_{\es}$ [Fig. \ref{Fig1}b], which gives
\begin{equation}
\omega_{\text{NV}} = \sqrt{\overline{\Omega}_{\es}^2 + (\omega_{\text{MW}} - \omega_{\es})^2} 
\label{Eqn2}
\end{equation}
leading to maximal probability oscillation between the $\ket{0}$ and $\ket{-1}$ states of the NV spin. Note that there exists an optimal EMR driving condition when the environmental spins are driven resonantly, $\omega_{\mw}=\omega_{\es}$, maximising the resulting NV Rabi frequency, $\Omega_{\nv}$, within the matching condition, Eq. (\ref{Eqn2}).

\paragraph*{Experiment -}

EMR driving of a single NV centre by an ensemble of environmental electron spins is achieved using an electronic-grade diamond crystal with $^{15}$NV centres implanted $5$-$15$ nm below the surface \cite{SI}. Free-electron spins known to exist at the diamond surface comprise the environmental spin ensemble [Fig. \ref{Fig1}a] \cite{Mamin2012a,Rosskopf2014,Myers2014,Sushkov2014b,Romach2015}. Due to the $2.87$ GHz zero-field splitting between the $\ket{0}$ and $\ket{-1}$ NV spin-states, and experimental difficulties in achieving GHz Rabi frequencies of the environmental spins, the EMR matching condition is most conveniently achieved near the ground-state level anti-crossing (GSLAC), which occurs at an external field strength $B_0=1024$ G [Fig. \ref{Fig1}c]. These experiments are performed at external field strengths giving $\omega_{\text{NV}}$ in the range $0$-$10$ MHz, and $\omega_{\es}$ in the range $2860$-$2880$ MHz accordingly.

In this regime, the NV electronic spin-state structure is complicated by hyperfine interaction with the intrinsic nuclear spin of the NV, giving rise to multiple hyperfine shifted transitions \cite{He1993,Broadway2016}. To simplify the EMR resonance landscape, an external field of $B_0 \approx 1023$ G was chosen, giving a single degenerate NV transition frequency at $\omega_{\text{NV}}\approx2.65$ MHz, as determined by directly driven magnetic resonance [Fig. \ref{Fig2}a]. Here the photo-luminescence (PL)  is a measure of the population of the $\ket{0}$ state, and the decrease on-resonance indicates driving of the $\ket{0}$ to $\ket{-1}$ transition when $\omega_{\text{MW}}=\omega_{\text{NV}}$.

To measure the EMR landscape, we use the pulse sequence illustrated in Fig. \ref{Fig2}b, where the microwave pulse frequency, $\omega_{\mw}$, is swept across the environmental spin transistion frequency, $\omega_{\es}$, and the driving field amplitude, $\overline{\Omega}_{\es}=\gamma_e B_1/2$ \cite{SI}, is swept across the NV transition, $\omega_{\nv}$. The microwave pulse duration, $\tau$, is fixed to maximise PL contrast at the optimal EMR driving condition, and a single laser pulse is used for optical readout and re-pumping of the NV spin-state. Figures \ref{Fig2}c and \ref{Fig2}d show EMR PL maps as a function of $\omega_{\text{MW}}$ and $\overline{\Omega}_{\es}$ as measured in experiment and predicted by Eq. (\ref{Eqn1}) respectively. The dashed lines show the EMR matching condition given by Eq. (\ref{Eqn2}), centred about the optimal driving condition where $\omega_{\text{MW}}=\omega_{\es}=2868$ MHz, and $\overline{\Omega}_{\es}=\omega_{\text{NV}}=2.65$ MHz. The resonant branches emanating from this point arise from the ability to recover the matching condition, $\Omega_{\es}=\omega_{\text{NV}}$, when the environmental spins are driven off-resonance, $\omega_{\text{MW}}\neq\omega_{\es}$, by reducing the driving strength, $\overline{\Omega}_{\es}$. The experimental data is found to be in good overall agreement with the theoretical model, as indicated by the line cuts presented in Figs. \ref{Fig2}e and \ref{Fig2}f. The broadening of the experimental map in driving strength as compared to the theoretical plot is attributed to magnet drift throughout the acquisition time (10 hours for Fig. \ref{Fig2}c).

\begin{figure}[t]
	\begin{center}
		\includegraphics[width=0.42\textwidth]{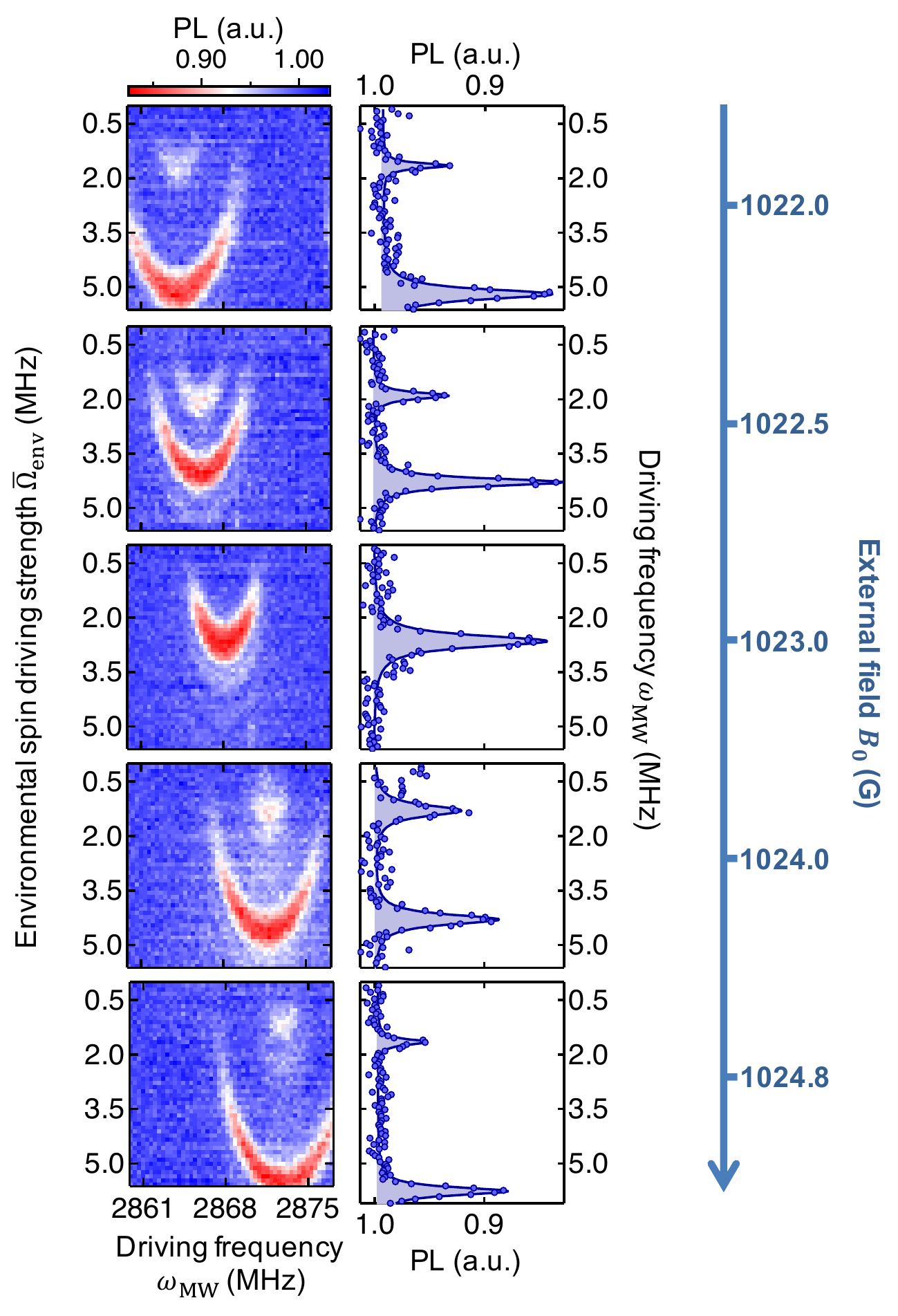}
		\caption{EMR PL maps (left) measured at external field strengths, $B_0$, in the range $1022.0$ - $1024.8$ G, increasing down the page, across the $^{15}$NV hyperfine crossing. Corresponding directly driven magnetic resonance spectra are given (right), identifying a distinct resonance feature in the EMR map for each of the hyperfine transitions. EMR map resonance features shift to higher driving frequency as the external field is increases in accordance with the environmental spin Zeeman splitting illustrated in Fig. \ref{Fig1}c.}
		\label{Fig3}
	\end{center}
\end{figure}

In general there are two $^{15}$NV hyperfine transitions about the GSLAC, which overlap at $B_0=1023$ G \cite{Broadway2016}. Repeating the previous measurement at various external field strengths across the GSLAC reveals this hyperfine structure as multiple resonance features in the EMR resonance landscape [Fig. \ref{Fig3}]. These EMR features match, in terms of $\overline{\Omega}_{\nv}$, the hyperfine transitions resolved by direct driving of the NV (right-hand side in Fig. \ref{Fig3}), with matching PL contrasts. This demonstrates the ability to selectively drive NV hyperfine transitions with EMR driving, by virtue of the relatively low power of the local driving fields involved. In addition, the centre of these resonance features shifts in driving frequency with increasing external field strength, in accordance with the Zeeman splitting of the free-electron spin-states, $\omega_{\es}=\gamma_e B_0$ (see Fig. \ref{Fig1}c).

\begin{figure}[t]
	\begin{center}
		\includegraphics[width=0.5\textwidth]{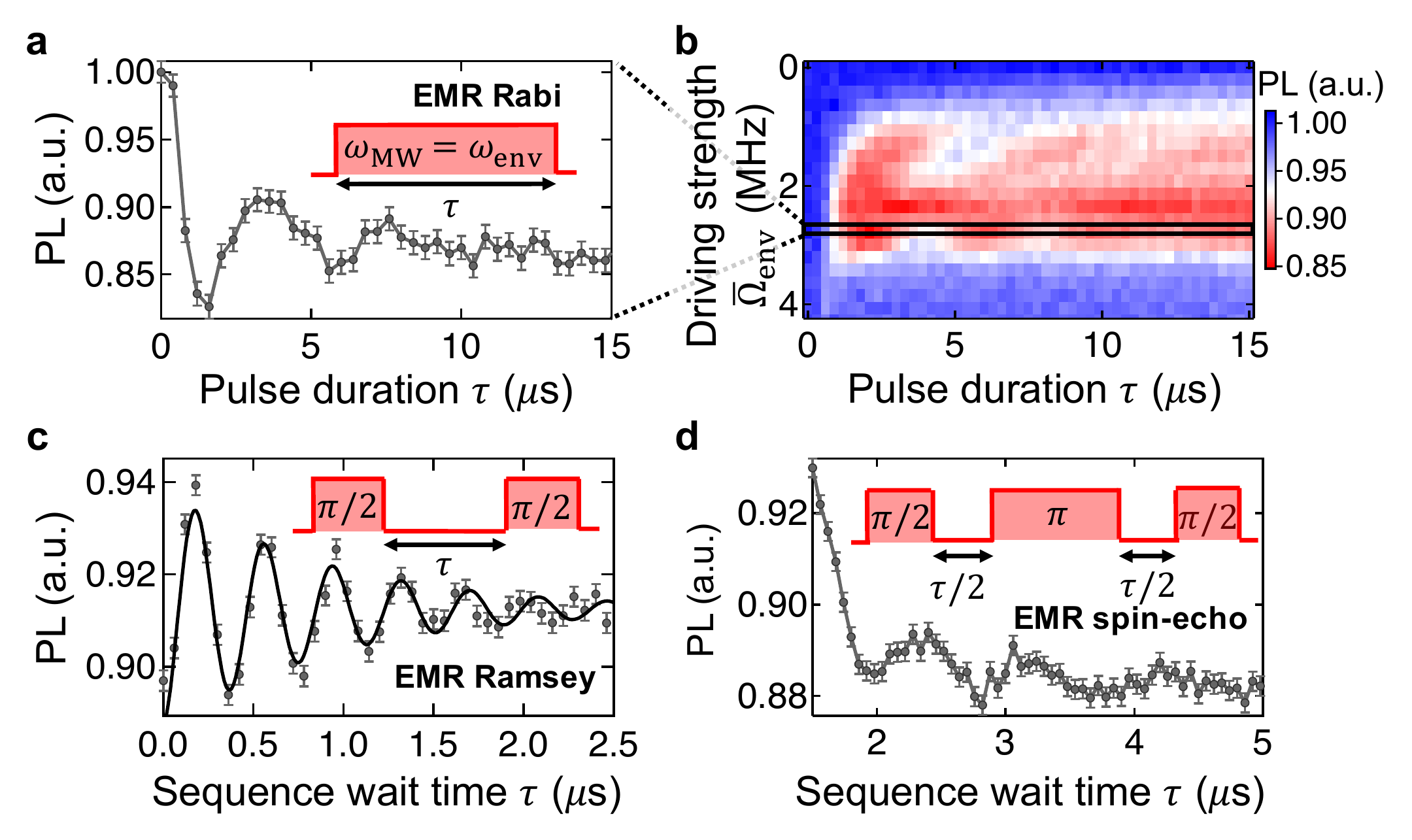}
		\caption{\textbf{a}, Optimally driven EMR Rabi curve with $\omega_{\text{MW}}=\omega_{\es}$ and $\Omega_{\es}=\omega_{\text{NV}}$ at the $^{15}$NV hyperfine crossing, $\omega_{\text{NV}}=2.65$ MHz, with microwave pulse scheme as inset. Effective $\pi/2$- and $\pi$-pulse times are identified at $650$ ns and $1300$ ns respectively. \textbf{b}, EMR driven Rabi PL map as a function of driving pulse duration and driving strength. Optimally driven Rabi curve using parameters identical to Fig. \ref{Fig4}a is highlighted. A driving frequency $\omega_{\mw}=2868$ MHz is used for driving strengths $\overline{\Omega}_{\es}>2.65$ MHz, and $\omega_{\mw}$ is reduced for $\overline{\Omega}_{\es}<2.65$ MHz, such that the EMR matching condition is satisfied. \textbf{c}, EMR driven Ramsey measurement using the optimal driving parameters and identified $\pi/2$-pulse durations of Fig. \ref{Fig4}a, with microwave pulse sequence as inset. The NV free induction decay shows a characteristic oscillation at the NV transition frequency, $2.65$ MHz, as determined by fit (solid line). \textbf{d}, Optimally driven EMR spin-echo curve using pulse durations identified in Fig. \ref{Fig4}a, with microwave pulse sequence as inset.}
		\label{Fig4}
	\end{center}
\end{figure}

These measurements, which utilised a fixed driving pulse duration maximising PL contrast, demonstrate the ability to induce spin-transitions of a target NV centre by EMR driving. We now probe the EMR driving dynamics by time resolved measurements, allowing the coherence of the observed control to be assessed. Utilising the optimal driving parameters, $\omega_{\text{MW}}=\omega_{\es}$ and $\overline{\Omega}_{\es}=\omega_{\text{NV}}$, identified in Fig. \ref{Fig2}c at the $^{15}$NV hyperfine crossing, an EMR driven Rabi curve on the NV was measured by varying the driving pulse duration [Fig. \ref{Fig4}a]. An oscillation with a period of $3\,\mu$s is observed, demonstrating coherent control of the NV spin-state. The corresponding Rabi frequency, $\Omega_{\text{NV}}=3.3 \,\text{MHz}\,=\gamma_e\,\alpha(\{\vec{r}_i\})$, is intrinsically linked to the spatial distribution of environmental spins, offering a pathway towards spatial mapping of such spins with nanoscale resolution. The rapidly decaying envelope of the Rabi curve arises as a consequence of the random initial spin-state of the environmental spin ensemble, such that the resulting curve is an average across a distribution of effective driving strengths of the environmental spin field \cite{SI}. 

We note that EMR Rabi driving can be achieved for any pair of driving parameters, $(\omega_{\text{MW}},\overline{\Omega}_{\es})$, satisfying the EMR matching condition depicted by the dashed line in Fig. \ref{Fig2}c. This is illustrated in Fig. \ref{Fig4}b, showing EMR Rabi curves as a function of $\overline{\Omega}_{\es}$, with the driving frequency, $\omega_{\text{MW}}$, chosen such that the EMR matching condition is satisfied where possible. When $\overline{\Omega}_{\es}>\omega_{\text{NV}}$, the driving frequency is fixed at $\omega_{\text{MW}}=\omega_{\es}$, as the EMR matching condition cannot be recovered in this regime, resulting in a sharp decrease in PL contrast (lower half Fig. \ref{Fig4}b). Decreasing the driving strength below the optimal condition, $\overline{\Omega}_{\es}<\omega_{\text{NV}}$, preserves the contrast, but gives a longer Rabi period according to the factor $(\overline{\Omega}_{\es}/\Omega_{\es})^2$ (upper half Fig. \ref{Fig4}b).

The coherent control demonstrated in Figs. \ref{Fig4}a and \ref{Fig4}b suggests the feasibility of using EMR to drive pulsed quantum control schemes fundamental to quantum information and quantum sensing protocols. Identifying effective $\pi/2$- and $\pi$-pulse durations from the optimally driven EMR Rabi curve at $650$ ns and $1300$ ns respectively, Ramsey and spin-echo measurements were performed. The free induction decay curve as measured by the EMR driven Ramsey sequence [Fig. \ref{Fig4}c] shows an oscillation at approximately $2.65$ MHz, the NV transition frequency. This oscillation arises from the phase accumulation of the NV spin-state relative to the effective driving field of the environmental spin ensemble, whose phase is effectively frozen during the free evolution time \cite{SI}. An analytic treatment in the macrospin approximation of the NV state evolution under this driving scheme reveals this oscillation, giving the probability of measuring the $\ket{0}$ state as $P_{\ket{0}}(\tau) = \left(1-\cos(\omega_{\nv}\tau)\right)/2$, where $\tau$ is the free evolution time of the sequence \cite{SI}. This feature is in contrast with directly driven Ramsey measurements, which exhibit an oscillation at the detuning frequency \cite{Maze2012}. The spin-echo sequence, by design, filters out effects from quasi-static dephasing processes \cite{SI}. Consequently, the EMR driven spin-echo curve [Fig. \ref{Fig4}d] shows revivals at a frequency of $1.1$ MHz, corresponding to the Larmor precession of the surrounding bath of $^{13}$C nuclear spins \cite{Childress2006}. In addition, theoretical analysis shows that the decay of the Ramsey and spin-echo measurements are dominated by the decoherence of the NV, with characteristic time scales $T_2^*$ and $T_2$ respectively, and the randomised initial states and decoherence of the environmental spins result primarily in a reduced constrast \cite{SI}.

\begin{figure}[t]
	\begin{center}
		\includegraphics[width=0.95\columnwidth]{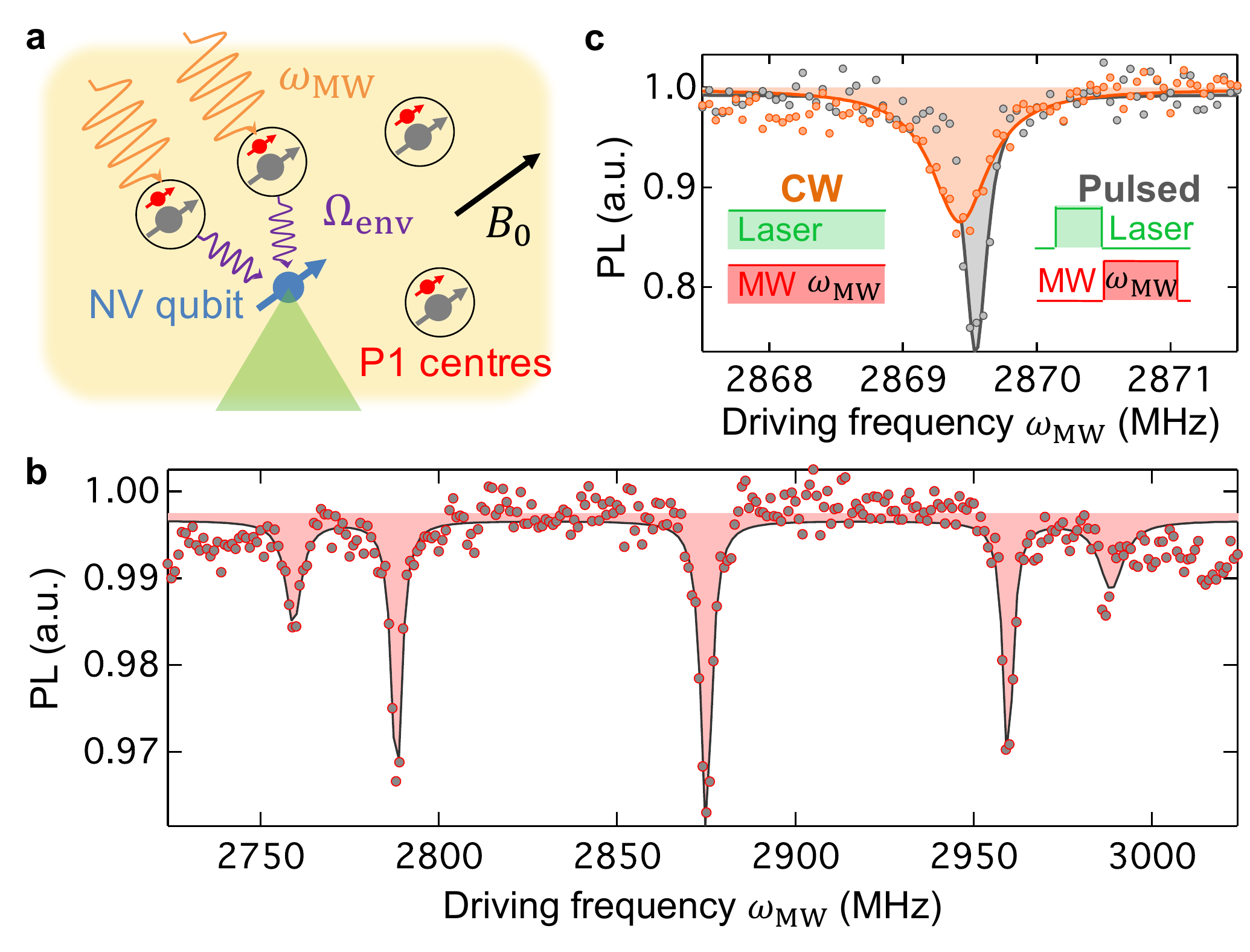}
		\caption{\textbf{a}, EMR schematic for driving of substitutional nitrogen (P1) centre electron spins (red) in bulk diamond. P1 centre nuclear spins (grey) lead to hyperfine splitting of the electron spin transition. \textbf{b}, Characteristic P1 centre spectrum acquired by EMR spectroscopy at $B_0\approx1024$ G. The FWHM of central resonance peak is approximately $5$ MHz. \textbf{c}, Continuous wave (CW) and pulsed EMR spectra of near surface free-electron species, with control schemes given as insets. Offset between the spectra is due to a variation in the external field strength used for each measurement.}
		\label{Fig5}
	\end{center}
\end{figure}

As a final experiment, we illustrate the applicability of EMR to spectroscopy by acquiring an electron spin resonance spectrum of a non-trivial spin species, namely substitutional nitrogen (P1) centres internal to a nitrogen rich host diamond [Fig. \ref{Fig5}a]. A representative P1 spectrum acquired by EMR driving is given in Fig. \ref{Fig5}b, revealing the characteristic five-peak structure of the centre due to the on- and off-axis parallel hyperfine interaction between the spin-1 $^{14}$N nuclear spin and spin-1/2 electron spin of the centre \cite{Smith1959,Hall2016,Wood2016}. We note that the resonance line width, which sets the spectral resolution of the technique, is governed by $\omega_{\text{NV}}$ for a given environmental spin ensemble \cite{SI}, and can therefore be improved by using an $^{14}$NV centre, for which $\omega_{\text{NV}}$ can typically be reduced to $100$ kHz \cite{Broadway2016}.

Applications of EMR to spectroscopy are made particularly attractive due to its experimental simplicity as compared to competing techniques, such as double electron-electron resonance (DEER) \cite{SI}, which requires pulsed driving of the qubit probe and environmental spin species in parallel \cite{Grotz2011,Mamin2012a}, and $T_1$-based spectroscopy, which requires delicate control of the varying magnetic field \cite{Hall2016,Wood2016}. The EMR protocol can be further simplified by implementing a continuous wave (CW) optical and microwave excitation scheme, achieving similar results as compared to the pulsed scheme [Fig. \ref{Fig5}b]. The reduced fluorescence contrast of the CW scheme is ascribed to the continuous optical re-pumping of the NV spin-state \cite{Manson2006}.

In this paper we have introduced a technique by which the coherent driving of a qubit is achieved by using nearby environmental spins as agents of control. This concept has been realised in experiments using a single NV centre in diamond, driven by an ensemble of electron spins, both at the diamond surface and in the bulk. The parameter space of this technique has been explored and compared to the simple semi-classical model developed, showing good agreement. Applications to spectroscopy have been demonstrated by acquisition of a characteristic substitutional nitrogen centre spectrum, and are made attractive by the experimental simplicity of the technique. Finally, the highly localised driving fields utilised by EMR provide an avenue by which target qubits within an array can be selectively addressed, especially in conjunction with environmental spin engineering.
 
The authors acknowledge useful discussions with L.T. Hall. This work was supported in part by the Australian Research Council (ARC) under the Centre of Excellence scheme (project No. CE110001027), and by the Melbourne Centre for Nanofabrication (MCN) in the Victorian Node of the Australian National Fabrication Facility (ANFF). L.C.L.H. acknowledges the support of an ARC Laureate Fellowship (project No. FL130100119). J.-P.T acknowledges support from the ARC through the Discovery Early Career Researcher Award scheme (DE170100129) and the University of Melbourne through an Establishment Grant and an Early Career Researcher Grant. S.E.L and D.A.B are supported by an Australian Government Research Training Program Scholarship.
 
\bibliographystyle{apsrev4-1}
\bibliography{bib}

\vspace{2.5cm}

\begin{center}
{\Large Supplemental Material}
\end{center}

\setcounter{equation}{0}
\setcounter{figure}{0}
\setcounter{table}{0}
\makeatletter
\renewcommand{\theequation}{S\arabic{equation}}
\renewcommand{\thefigure}{S\arabic{figure}}
\renewcommand{\thesection}{S\arabic{section} }
\renewcommand{\thesubsection}{S\thesection.\arabic{subsection} }
\renewcommand{\thesubsubsection}{S\thesubsection.\arabic{subsubsection} }
	  
\section{Experiment details} \label{DiamondSample}

The diamond sample used for the experiments presented in Figs. $2$-$4$ and $5$c was a $[100]$ oriented, electronic grade CVD diamond crystal, grown on an HPHT bulk diamond substrate. NV centres were formed by ion implantation of $^{15}$N at an energy of $3.5$ keV, and a fluence of $10^9$ cm$^{-2}$, giving an expected implantation range of $5$-$15$ nm\cite{Wood2016a,Lehtinen2016}. The sample was annealed under UHV at $800$ C for $5$ hours. The resulting NV density was such that on average there was less than one centre per optically resolvable spot, approximately $(0.3\,\mu\text{m})^2$ within the shallow implantation sheet. The spin environment of the electronic grade CVD bulk resulted in typical NV centre coherence times of $T_2^*\approx3\,\mu$s. The diamond sample used for acquiring the P1 centre electron spin resonance spectrum [Fig. $5$b] was similarly fabirated, except for intentional doping with nitrogen during the CVD growth, ensuring a sufficiently high density of P1 centres to allow spectra to be measured by single NV centres.

The measurements were performed at room temperature using a home-built confocal microscope similar to that described in \cite{Wood2016}. The external magnetic field was applied using a permanent magnet and aligned with the NV axis by maximising the PL intensity \cite{Wood2016}.

\section{Driving strength calibration} \label{CalibES}

The EMR data presented gives the direct driving field amplitude in terms of the driving strength of the environmental spins proximal to the NV, $\overline{\Omega}_{\es}$, which is proportional to the local amplitude of the applied microwave field, $B_1$, according to the relation $\overline{\Omega}_{\es}=\gamma_e B_1/2$. Experimentally, the calibration of $B_1$, and hence $\overline{\Omega}_{\es}$, was achieved for a given environmental spin species at a target NV site in the following manner. The target NV centre was driven resonantly, $\omega_{\mw}=\omega_{\nv}$, at a range of driving field amplitudes, $B_1$, and Rabi measurements made for both the $\ket{0}\leftrightarrow\ket{-1}$ and $\ket{0}\leftrightarrow\ket{+1}$ NV transitions. These measurements were performed at a low external field strength, $B_0=5$ G, such that the two NV transition were resolvable, but close to the expected environmental spin transition frequency under EMR conditions, $\omega_{\es}\approx2870$ MHz. Rabi frequencies were extracted by fitting the measured Rabi curves, and averaged between the two transitions for a given driving field amplitude, mitigating any frequency dependence in driving strength. The equivalent Rabi frequency for a spin-1/2 environmental electron system was calculated by scaling the transition averaged NV Rabi frequencies by $\sqrt{2}$, due to the differing pre-factor in the definition of the spin matrices for the spin-1/2 system versus the spin-1 NV system.

\begin{figure}[b]
	\begin{center}
		\includegraphics[width=0.35\textwidth]{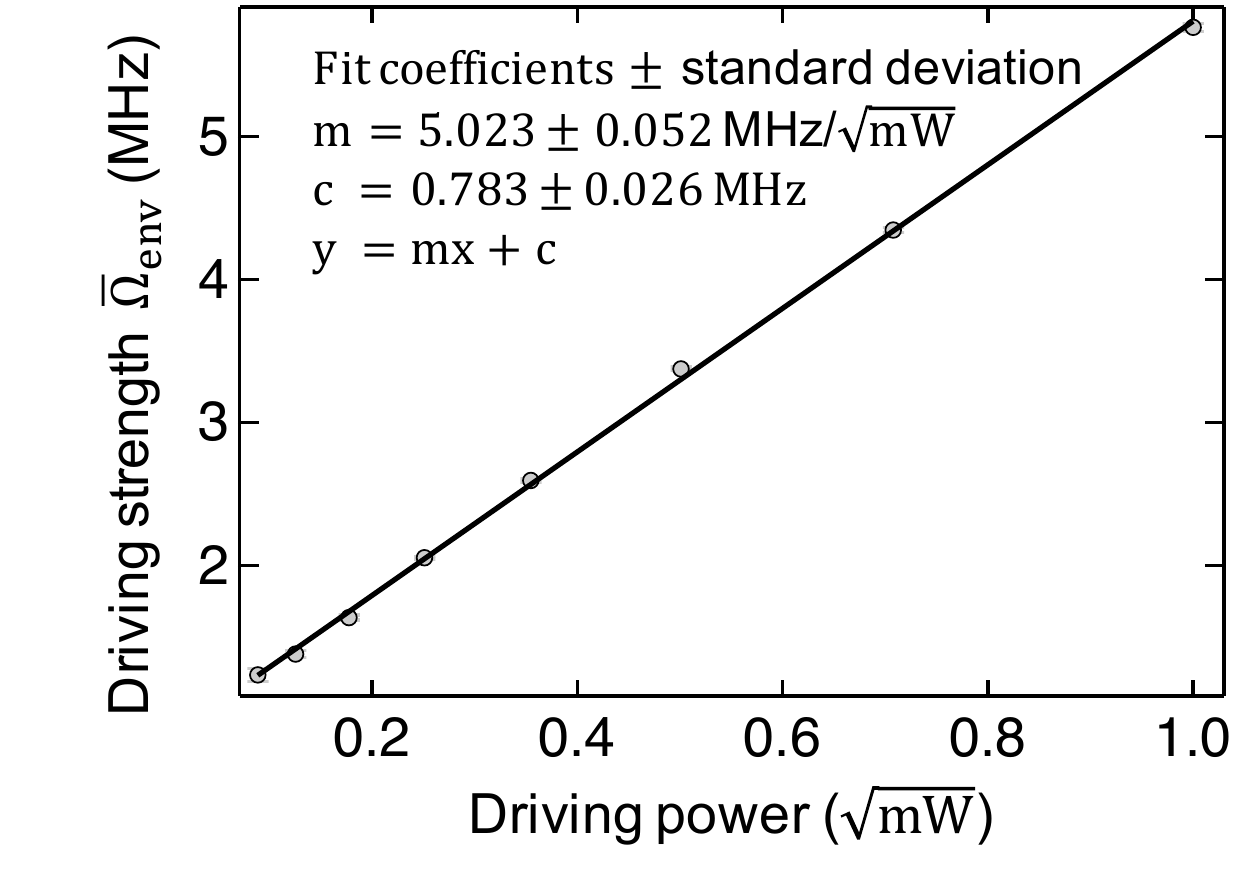}
		\caption{Environmental spin driving strength calibration curve as a function of driving field power in $\sqrt{\text{mW}}$, calculated for the single NV centre studied in Figs. $2$-$4$.}
		\label{FigS1}
	\end{center}
\end{figure}

Fig. \ref{FigS1} shows the environmental spin driving strength calculated in this fashion as a function of the driving field amplitude, here given as the square root of the driving power. The linear fit is the calibration curve used to calculate the driving strength of spin-1/2 electron spins in the environment of the target NV for the subsequent EMR experiments. It should be noted that this calibration is specific to environmental spins in close proximity to the target NV, given the spatial dependence of the magnetic field emanating from the micro-wire driving the bath.

\section{Single environmental macrospin model} \label{RabiModel}

The semi-classical Rabi formula model presented treats the environmental spin ensemble as a single macrospin. For simplicity, the spin-1/2 macrospin with transition frequency $\omega_{\es}$ is treated, in the first instance, as being initialised in the spin-down state, $\ket{\psi(t=0)}=\ket{\downarrow}$, and is directly driven by an oscillating magnetic field of the form
\begin{equation}
\vec{B}_{\mw}(t)=B_1\cos(\omega_{\mw} t)\,\hat{e}_\perp \label{DirDriv}
\end{equation}
where $B_1$ and $\omega_{MW}$ are the direct driving field amplitude and frequency respectively, and $\hat{e}_\perp$ denotes the vector component perpendicular to the macrospin quantisation axis. Applying Rabi's formula in this context gives the probability of measuring the macrospin in the $\ket{\uparrow}$ state, after a driving time $\tau$ as
\begin{equation}
P_{\ket{\uparrow}}(\tau) = \left(\dfrac{\overline{\Omega}_{\es}}{\Omega_{\es}}\right)^2 \dfrac{1 - \cos(\Omega_{\es} \tau)}{2} \label{MacroProp}
\end{equation}
where $\Omega_{\es}= \sqrt{\overline{\Omega}_{\es}^2 + (\omega_{\text{MW}} - \omega_{\es})^2}$, is the environmental spin Rabi frequency with driving strength $\overline{\Omega}_{\es}=\gamma_eB_{1}/2$, where $\gamma_e$ is the electron gyromagnetic ratio. 

The effect of the environmental ensemble spin-state oscillations on a proximal NV centre is modelled semi-classically, treating the effective field arising from the driven environmental spins as a classical field driving the NV centre. In the initialised macrospin model, this field can be expressed simply as
\begin{equation}
 \vec{B}_{\es}(\tau)=B_{\es}\cos(\Omega_{\es}\tau)\hat{e}_{\perp} \label{FieldES}
\end{equation}
where $B_{\es}=\gamma_e\,\alpha(\{\vec{r}_i\})\, (\overline{\Omega}_{\es}/\Omega_{\es})^2$ is the amplitude of the field arising from the environmental spin oscillations perpendicular to the NV-axis. The factor $\alpha(\{\vec{r}_i\})$ accounts for the dependence of the environmental spin field on the relative position of the environmental spins and the NV, $\vec{r}_i$, and can be extended to encapsulate the dependence on the initial states of the spins comprising the ensemble (see section \ref{Ensemble}). Applying the Rabi formula to an NV initialised in the $\ket{0}$ state driven by this field, Eq. ($1$) is recovered, where only transitions to the $\ket{-1}$ state are considered due to limitations in the achievable range of environmental spin Rabi frequencies.

This semi-classical model is validated by the experimental observation of matching PL contrast between optimal EMR driving, centre of Fig. $2$c, and direct driving of the NV transition, Fig. $2$a. Back action effects present in a fully quantum model would reduce the observed PL contrast in EMR driving.

\section{Extension to environmental spin ensemble} \label{Ensemble}

The macrospin model presented ignores the details of individual spins comprising the ensemble, and consequently offers no explanation for the observed rapid decay of the EMR driven Rabi oscillations [Fig. 3a]. Here we extend the model to consider both the position and intial states of the environmental spins comprising the ensemble, and attempt to explain such details. 

The magnetisation of a single spin-1/2 electron spin along is quantisation axis, $\hat{e}_z$, driven by a direct driving field of the form Eq. (\ref{DirDriv}), from an arbitrary initial state, $\ket{\psi_{\es}(t=0)} = \cos(\theta_{\es})\ket{\downarrow} \, + \, \exp(i\phi_{\es}) \sin(\theta_{\es}) \ket{\uparrow}$, can be expressed as
\begin{align}
\vec{m}_{\es}(t) &= \mu_B \left[\cos(2\theta_{\es})\cos(\Omega_{\es}t) \right. \nonumber \\ 
 & \qquad + \left.\sin(\phi_{\es})\sin(2\theta_{\es})\sin(\Omega_{\es}t)\right] \, \hat{e}_z 
\label{MagES}
\end{align}
where $\theta_{\es}$ and $\phi_{\es}$ are respectively the polar and azimuthal angles of the initial state vector on the environmental spin Bloch sphere, and $\mu_B$ is the Bohr magneton. Note that the environmental spin quantisation axis is defined by the external magnetic field used to Zeeman split the bath spin-states, and is therefore aligned with the NV-axis. 

The magnetic field arising from this magnetisation at a proximal NV centre, perpendicular to the NV quantisation-axis, is given by
\begin{align}
\vec{B}_{\es}^{\perp}(t) &= \dfrac{3\mu_0\mu_B}{8\pi} \left[\left(E\cos(\Omega_{\es}t) + F\sin(\Omega_{\es}t)\right)\hat{e}_x \right. \nonumber \\
& \qquad \left.+ \left(G\cos(\Omega_{\es}t) + H\sin(\Omega_{\es}t)\right)\hat{e}_y\right] \label{Bestot}
\end{align}
where $E$, $F$, $G$, and $H$ encapsulate the initial state and position dependence, and are defined as
\begin{align}
E &= \sin(2\theta)\cos(\phi) \cos(2\theta_{\es})/r^3 \\
F &= \sin(2\theta)\cos(\phi) \sin(\phi_{\es}) \sin(2\theta_{\es})/r^3 \\
G &= \sin(2\theta)\sin(\phi) \cos(2\theta_{\es})/r^3 \\
H &= \sin(2\theta)\sin(\phi) \sin(\phi_{\es})\sin(2\theta_{\es})/r^3
\end{align}
and the environmental spin position relative to the NV is given by $\vec{r}=(r,\theta,\phi)$. The above expression is extended to environmental ensemble of $n$ spins by summing over the individual spin positions, which are fixed in relation to the NV, and their initial states, which will vary randomly between measurements due to spin-lattice relaxation. Performing this summation and comparing the effective field to a direct driving field, it is possible to define the driving strength of the NV by this field as 
\begin{align}
\overline{\Omega}_{NV} = \dfrac{3\mu_0\mu_B\gamma_e}{8\sqrt{2}\pi} \, \left|E - iG + iF + H\right|
\end{align}
where $E$, $F$, $G$, and $H$ are the previous quantities summed over the ensemble.

\begin{figure*}[t]
	\begin{center}
		\includegraphics[width=0.9\textwidth]{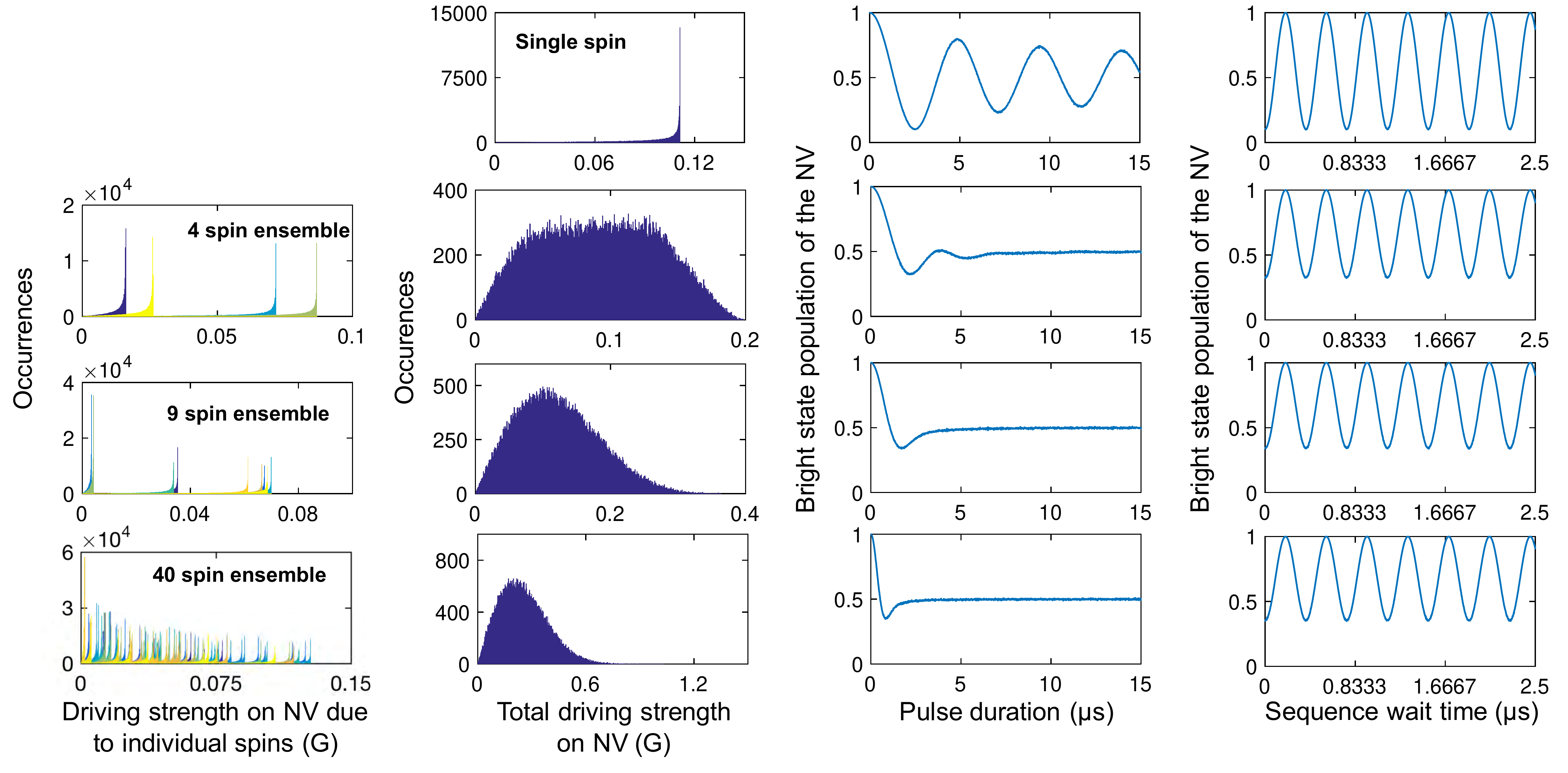}
		\caption{Numerical simulations of EMR driving of a single NV centre by randomly initialised environmental spin ensembles of varying sizes. The first column shows the driving strength distributions of the individual spins comprising the ensemble given a fixed position but varying initial spin state. The second column shows the corresponding total driving strength distribution of the ensemble on the NV. The third and fourth columns shows the resulting EMR driven Rabi and Ramsey curves respectively, averaged across $10000$ initial ensemble states at each time point.}
		\label{FigS2}
	\end{center}
\end{figure*}

Having developed this formalism, numerical simulations can be performed to determine the effect of the randomised initial state of the environmental spin ensemble on the EMR driving of the NV. Environmental spin ensembles were created by randomly assigning spin positions within a $5$-$6$ nm shell centred on a single NV centre, with the initial state of each spin chosen from a uniform distribution about their respective Bloch spheres. Keeping the position of each spin within the ensemble fixed, but varying the initial states at each measurement, the driving strength distribution of the ensemble can be produced, and the resulting EMR driven Rabi and Ramsey curves of the NV calculated using Eq. ($1$) and Eq. (\ref{EMRonresRamsey}) respectively.

Figure \ref{FigS2} presents numerical simulation data for four environmental spin ensembles, showing the driving strength distributions of each component spin (first column), the total driving strength distribution of the ensemble (second column), and demonstrative EMR driven Rabi and Ramsey curves of the proximal NV (third and fourth columns respectively). The characteristic shape of the individual spin distributions arises from the random initial state given by Eq. (\ref{MagES}), whereas the total driving strength distribution depends on the spatial arrangement of the environmental spins relative to both the NV, and the other environmental spins. Broadly, a greater mean driving strength of the ensemble results in a higher EMR Rabi frequency, and the breadth of the distribution governs the decay of the Rabi envelope. The $\pi/2$-pulse durations used to model the Ramsey measurements are taken from the corresponding Rabi curve, and hence the minimum population observed in Rabi is commensurate with the initial population in the Ramsey measurement. Note that this model ignores decoherence and detuning effects on both the NV and the environmental spins. These effects are discussed in section \ref{Detuning}.

\section{Ramsey and spin-echo sequence modelling} \label{Analytics}
 
 The oscillation observed in the free induction decay curve of the EMR driven Ramsey sequence [Fig. $4$c], can be explained by analytic treatment of the NV spin-state evolution driven by the effective field arising from the environmental spin ensemble. The treatment developed in section \ref{Ensemble} allows the environmental spin field perpendicular to the NV axis to be expressed as 
 \begin{align}
 \vec{B}_{\es}(t) = B_1 \cos(\Omega_{\es} t - \Phi)\hat{e}_\perp
 \end{align}
 where $B_1=\dfrac{3\mu_0\mu_B}{8\pi}\left|E - iG + iF + H\right|$, and $\Phi$ is an arbitrary phase of the field due to the initial ensemble state. Treating this field as a classical field driving the an initialised NV resonantly, $\Omega_{\es}=\omega_{NV}$, the evolution of the NV state can be found by solving the time-dependent Schr\"odinger equation, neglecting decoherence effects for the moment. Note that this driving field does not accumulate phase during the sequence wait time, given that its phase arises from the longitudinal projections of the ensemble spin states, which do not evolve when the direct driving field is switched off.
 
 Solving for the NV state evolution due to the EMR driven Ramsey sequence as outlined, the final NV state is given by
 \begin{align}
 \ket{\psi_{\nv}(t_3)} =& \, i \sin\left(\dfrac{\omega_{\nv}}{2}\tau\right) e^{+i\omega_{\nv} t_1} \ket{0}  \nonumber \\
 &- i \cos\left(\dfrac{\omega_{\nv}}{2}\tau\right) e^{-i(\omega_{\nv} t_1- \Phi)} \ket{-1}
 \end{align}
 where $t_3$ denotes the end of the sequence, $t_1= \pi/2\Omega_{\nv}$ is an idealised $\pi/2$-pulse duration, and $\tau$ is the sequence wait time. Performing a PL measurement of the above state reveals the observed oscillation at the NV transition frequency, $\omega_{\nv}$
 \begin{align}
 \left|\bra{0}\ket{\psi_{NV}(t_3)}\right|^2\,&=\, \dfrac{1-\cos(\omega_{NV}\tau)}{2}
 \end{align}
Additionally, this oscillation will decay over a time scale give by the NV decoherence time, $T_2^*$, due to fluctuations in $\omega_{\nv}$, as in tranditional Ramsey measurements. 
 
 The same analysis can be applied to the EMR driven spin-echo sequence to demonstrate its filtering of detuning-like effects arising from EMR driving, as seen in the Ramsey treatment. Using idealised $\pi/2$- and $\pi$-pulse durations as before, the NV state at the end of the spin-echo sequence is given by
 \begin{align}
 \ket{\psi_{\nv}(t_5)} &= - e^{i\omega_{\nv} (t_5-\tau)/2} \, \ket{0}
 \end{align}
 where $t_5$ is the total sequence duration, and $\tau$ is the total sequence wait time. Performing a PL measurement of this state will hence reveal the NV to be in the bright state
 \begin{align}
 \left|\bra{0}\ket{\psi_{\nv}(t_5)}\right|^2\,&= 1
 \end{align}
as in traditional spin-echo measurements, correcting for the detuning-like effects of EMR. Magnetic field oscillating synchronously with the spin-echo sequence lead to a reduced population of the $\ket{0}$ state and result in the modulation of spin-echo curves as the sequence wait time is increased, which otherwise show a simple exponential decay over the decoherence time of the NV, $T_2$.

\section{Decoherence effects in Ramsey and spin-echo}\label{Detuning} 
 
 The model developed in section \ref{Ensemble} incorporated both the positions and initial states of the environmental ensemble spins, and offered an explanation for the observed decay in EMR driven Rabi oscillations, and a corresponding reduction in constrast of the EMR driven Ramsey measurement. Here we make explicit the form used for the previous Ramsey simulations, which builds on analytic framework outlined in section \ref{Analytics}.
 
Neglecting decoherence and detuning effects, the NV state at the end of the Ramsey sequence, $t=t_3$, is given by
 \begin{widetext}
 \begin{align}
 	\ket{\psi_{\nv}(t_3)} =& \, \left[\cos^2\left(\dfrac{\Omega_{\nv}}{2}t_1\right) e^{+i\omega_{\nv} \tau/2} -\sin^2\left(\dfrac{\Omega_{\nv}}{2}t_1\right) e^{-i\omega_{\nv} \tau/2}\right] e^{+i\omega_{\nv} t_1} \ket{0}  \nonumber \\
 	& - i \sin\left(\Omega_{\nv}t_1\right)\cos\left(\dfrac{\omega_{\nv}}{2}\tau\right) e^{-i(\omega_{\nv} t_1+ \Phi)} \ket{-1} \label{EMRonresRamsey}
 \end{align}
 \end{widetext}
 where $t_1$ is the chosen $\pi/2$-pulse duration used for the experiment and here $\Omega_{\nv}=\overline{\Omega}_{\nv}=\dfrac{3\mu_0\mu_B\gamma_e}{8\sqrt{2}\pi}\left|E - iG + iF + H\right|$, given that the EMR matching condition is satisfied. Calculating the probability of the NV ocupying the $\ket{0}$ state and averaging over the initial environmental spin states for a given ensemble, as in section \ref{Ensemble}, reveals sustained Ramsey oscillations (fourth column Fig. \ref{FigS2}) with reduced contrast as compared to the traditional Ramsey measurements, as discussed previously.
 
 \begin{figure*}[t]
 	\begin{center}
 		\includegraphics[width=\textwidth]{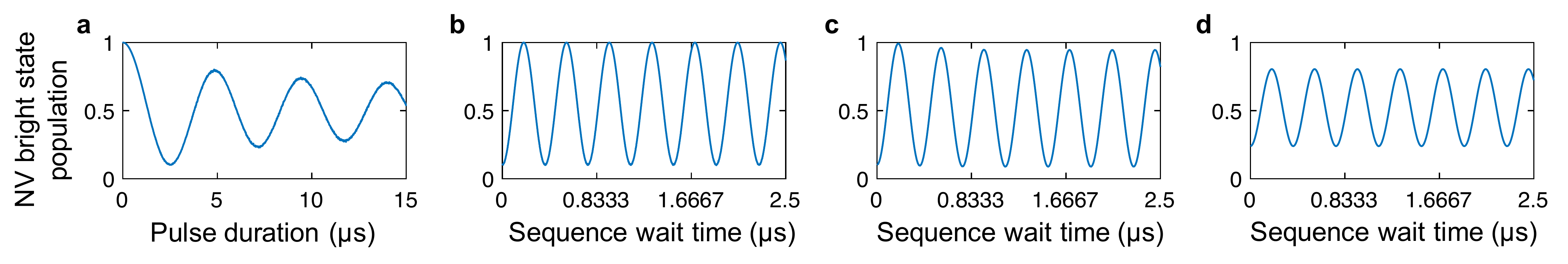}
 		\caption{\textbf{a}, Optimally driven NV Rabi curve due to a single macrospin at $r=5$ nm and $\theta = \pi/4$, relative to the NV, averaged over $10000$ initial macrospin states. $\pi/2$-pulse durations for subsequent Ramsey curves are extracted from this simulation. \textbf{b}, EMR Ramsey curve omitting decoherence and detuning effects, averaged over $10000$ initial macrospin states \textbf{c}, EMR Ramsey curve modelling environmental spin decoherence as quasistatic detuning, characterised by an environmental spin $T_2^*=1.0$ $\mu$s. Curve averaged over $1000$ initial macrospin states, and $1000$ quasistatic detunings per time point. \textbf{d} EMR Ramsey curve modelling environmental spin decoherence in the fast fluctuating limit. Curve averaged over $1000$ initial macrospin states, and $1000$ phase fluctuations per time point.}
 		\label{FigS5}
 	\end{center}
 \end{figure*}
 
 Having determined the effect of a random initial state of the environmental ensemble on the EMR driven Ramsey measurement, we now determine the effect of environmental spin decoherence throughout the free evolution time, $\tau$, in both quasistatic and fast fluctuating regimes. For simplicity, we again model the ensemble as a single macrospin. 
 
 Firstly, we model the environmental spin decoherence as arising from a quasistatic detuning between the driving field frequency, $\omega_{\mw}$, and the environmental spin transition frequency, $\omega_{\es}$, that is random but constant over the sequence duration. We consider an environmental macrospin with a general inital state 
 \begin{align}
 \ket{\psi_{\es}(t=0)} =\,& \cos(\theta_{\es})\ket{\downarrow} \,  \nonumber\\
  &+ \, \exp(i\phi_{\es}) \sin(\theta_{\es}) \ket{\uparrow}	
 \end{align}
driven by a direct driving field in the form of Eq. (\ref{DirDriv}), as before. The effective driving field on the NV arising from the directly driven macro spins is given by
\begin{align}
	\vec{B}_{\es}(t) = B_1 \cos(\Omega_{\es} t - \Phi)\hat{e}_\perp
\end{align}
 where the amplitude and initial phase of the field are given by
\begin{align}
	B_1 =\,& \dfrac{3\mu_0\mu_B\sin(2\theta)}{8\pi r^3} \nonumber \\
	&\,\,\,\times \sqrt{\cos(2\theta_{\es})^2 + \sin(\phi_{\es})^2\sin(2\theta_{\es})^2} \label{esamp} \\
    \Phi =\,& \atan(\sin(\phi_{\es})\tan(2\theta_{\es})) \label{esphi}
\end{align} 
respectively. The relative position of the macrospin and the NV is given by $(r,\theta)$.

Tracking the evolution of the macrospin and the NV throughout the sequence, a detuning term is introduced for the macrospin during the free evolution time, $\tau$. After the free evolution time, $t=t_2$, the macrospin state is given by 
 \begin{align}
	\ket{\psi_{\es}(t_2)} =& \, \alpha_2 \, e^{+i\delta_{\es}\tau/2} \ket{\downarrow} \, + \, \beta_2 \, e^{-i\delta_{\es}\tau/2} \ket{\uparrow}	
\end{align}
 where $\delta_{\es}=\omega_{\es}-\omega_{\mw}$ is the detuning, and the coefficients $\alpha_2$ and $\beta_2$ give the evolution of the initial macrospin state due to the first pulse. The detuning here is modelled as arising from magnetic flield fluctuations along the quantisation axis of the macrospin, $B_z$, with a standard deviation, $\sigma_{B_z}$, related to the decoherence time, $T_2^*$, of the macrospin as $\sigma_{B_z}=(\gamma_e T_2^*/\sqrt{2})^{-1}$.
 
 Evolving the macrospin following the free evolution time by application of the second $\pi/2$-pulse, the effective driving field on the NV due to the macrospin can be calculated, and applied to further evolve the NV, giving a final state
  \begin{widetext}
 	\begin{align}
 		\ket{\psi_{\nv}(t_3)} =& \, \left[\cos\left(\dfrac{\Omega_{\nv}}{2}t_1\right) \cos\left(\dfrac{\Omega_{\nv'}}{2}t_1\right) e^{+i\omega_{\nv} t_2/2} -e^{i(\Phi+\Phi')} \sin\left(\dfrac{\Omega_{\nv}}{2}t_1\right) \sin\left(\dfrac{\Omega_{\nv'}}{2}t_1\right) e^{-i\omega_{\nv} t/2}\right] e^{+i\omega_{\nv} t_1} \ket{0}  \nonumber \\
 		& - i e^{i \Phi} \left[\sin\left(\dfrac{\Omega_{\nv}}{2}t_1\right) \cos\left(\dfrac{\Omega_{\nv'}}{2}t_1\right) e^{-i\omega_{\nv} t_2/2} + e^{-i(\Phi+\Phi')} \cos\left(\dfrac{\Omega_{\nv}}{2}t_1\right) \sin\left(\dfrac{\Omega_{\nv'}}{2}t_1\right) e^{+i\omega_{\nv} t/2}\right] e^{-i\omega_{\nv} t_1} \ket{1} \label{DetNV}
 	\end{align}
 \end{widetext}
 where $\Omega_{\nv'}$ and $\Phi'$ are the driving strength and initial phase of the second marcospin field pulse after detuning throughout the free evolution time. Calculating $\left|\bra{0}\ket{\psi_{NV}(t_3)}\right|^2$ from Eq. (\ref{DetNV}) and numerically averaging of both the macrospin initial state and the macrospin detuning, the resulting Ramsey measurement simulated.
 
 Figure \ref{FigS5}c shows an EMR Ramsey curve with a quasitatic detuning characterised by $T_2^*=1.0$ $\mu$s, averaged over $1000$ detunings per time step and further averaged over $1000$ initial ensemble states. For comparison, the corresponding Ramsey curve omitting decoherence effects, but averaging over inital states for the same macrospin configurations is given [Fig. \ref{FigS5}b], along with its corresponding Rabi curve [Fig. \ref{FigS5}a]. It is clear that the quasistatic detuning results in a slight decay of the Ramsey envelope over the timescale given by $T_2^*$, but that the oscillations persist in the long term. 
 
 This quasistatic model can be extended to a fast fluctuation limit, where the dephasing of the macrospin throughout the free evolution time is modelled by reassigning the phase of macrospin state, $\phi_{\es}$, randomly from a uniform distribution on the interval $[0,2\pi]$, at the start of the second $\pi/2$-pulse of the sequence. The resulting driving field on the NV due to this random phase shift is calculated using Eqs. (\ref{esamp}) and (\ref{esphi}) and substituting into Eq. (\ref{DetNV}), from which the Ramsey measurement can be calculated as before.
 
 Figure \ref{FigS5}d shows the Ramsey measurement in this fast fluctuating limit for the same macrospin configuration as in Figs. \ref{FigS5}a, b and c, averaged over $1000$ phase fluctuations per time step, and $1000$ initial macrospin states. It is clear that the fast fluctuations further reduces the constrast of the Ramsey oscillations, a result supported by the low contrast Ramsey oscillations measured in experiment [Fig. $4$c].
 
 This analysis shows that the random initial state and decoherence effects of the ensemble limit the contrast of resulting EMR Ramsey oscillations, but do not dominate the decay of the Ramsey envelope. Consequently, the decay of the Ramsey envelope will be dictated by the decoherence of the NV itself, over the time scale given by $T_2^*$ of the NV. This reasoning is extrapolated to the case of the spin-echo sequence, where the decay timescale is instead characterised by $T_2$ of the NV.

 \section{P1 centre spectrum comparison with DEER} \label{P1DEER}
 
 The P1 centre spectrum presented in Fig. $5$b demonstrates the viability of EMR as a spectroscopic technique. Such a demonstration, however, is incomplete without comparison to established spectroscopic techniques. Here we present a P1 centre spectrum acquired using the same single NV centre as that of Fig. $5$c, but using double electron-electron resonance (DEER) instead of EMR.
 
\begin{figure}[t]
 	\begin{center}
 		\includegraphics[width=\columnwidth]{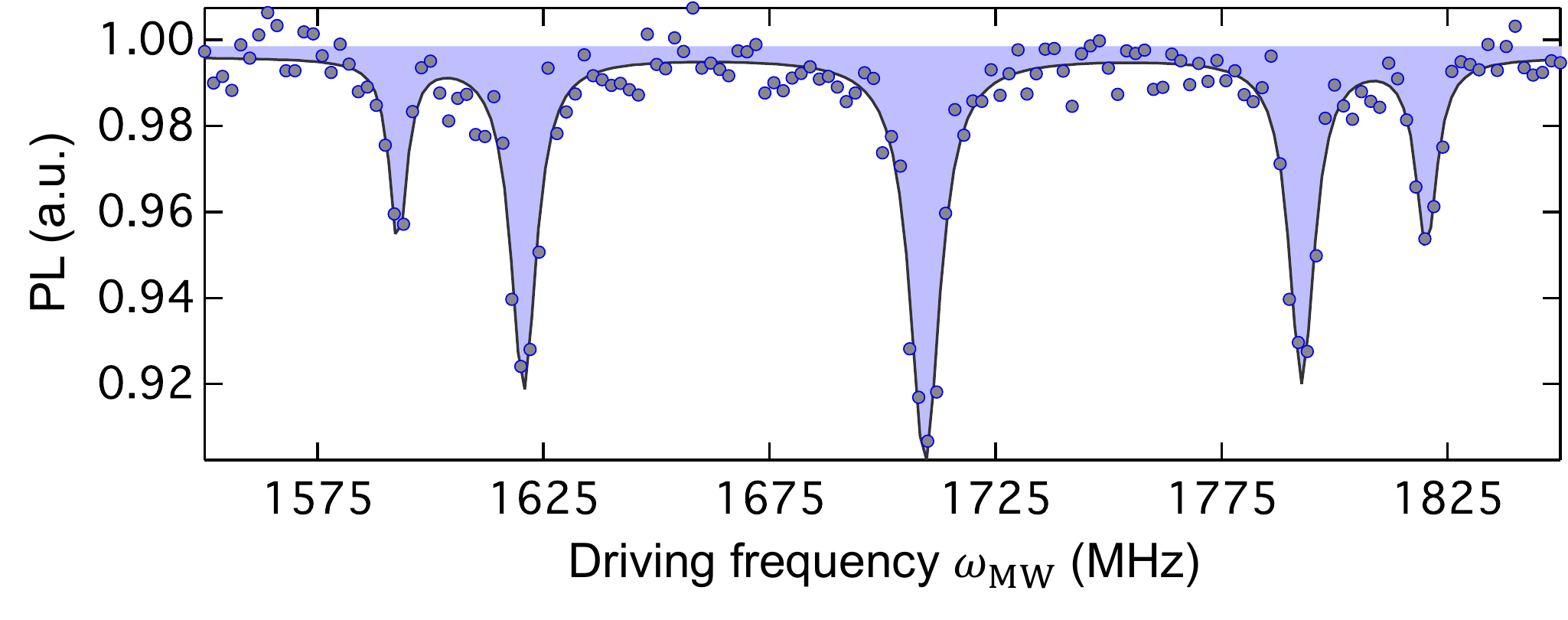}
 		\caption{Characteristic P1 centre spectrum acquired by DEER spectroscopy at $B_0\approx610$ G, where the FWHM of the central resonance peak is approximately $9$ MHz.}
 		\label{FigS4}
 	\end{center}
 \end{figure}

 The corresponding DEER spectrum is shown in Fig. \ref{FigS4}, where again, the characteristic five- peak structure arises from the on- and off-axis hyperfine interation between the centre's spin-1 $^{14}$N nuclear spin, and the spin-1/2 electron spin. The FWHM of the central resonance peak is approximately $9$ MHz in the DEER spectrum, as compared to the $5$ MHz in the EMR spectrum, which is attributed to the susceptibility of DEER to power broadening. 
 
 As a final note, a distinguishing feature of EMR spectroscopy is its probing of the $xz$ component of the dipole-dipole interaction between the NV and environmental spins, as compared to DEER and $T_1$ based spectroscopy, which probe the $zz$ and $xx$ terms respectively \cite{Grotz2011,Mamin2012a,Hall2016,Wood2016}. For this reason, EMR spectroscopy proves advantageous in detecting environmental spins in geometries to which established techniques are insensitive, such as the ``magic angle" in DEER \cite{Wood2016}. The combination of EMR with these established techniques thus provides a basis by which the spatial distribution of environmental spins can be mapped.
 
\section{Spectral resolution} \label{SpectralRes}
 
 The line width of resonance features in the EMR spectra limits the spectral resolution of the technique. This line width is naturally governed by the NV transition frequency, as this dictates the strength at which the environmental spin ensemble is driven, but is also governed by the effective field strength arising from the environmental spin ensemble at the site of the NV. The line width dependence on these parameters can be investigated numerically by solving Eq. ($1$) for the FWHM as a function of NV transition frequency, $\omega_{\nv}$, and driving strength of the NV by the ensemble field, $\overline{\Omega}_{\es}$. The driving pulse duration, $\tau$, is here chosen to be an idealised $\pi$-pulse for a given parameter set such that maximum PL contrast is observed, mimicking the experimental protocol.
 
  \begin{figure}[t]
 	\begin{center}
 		\includegraphics[width=0.36\textwidth]{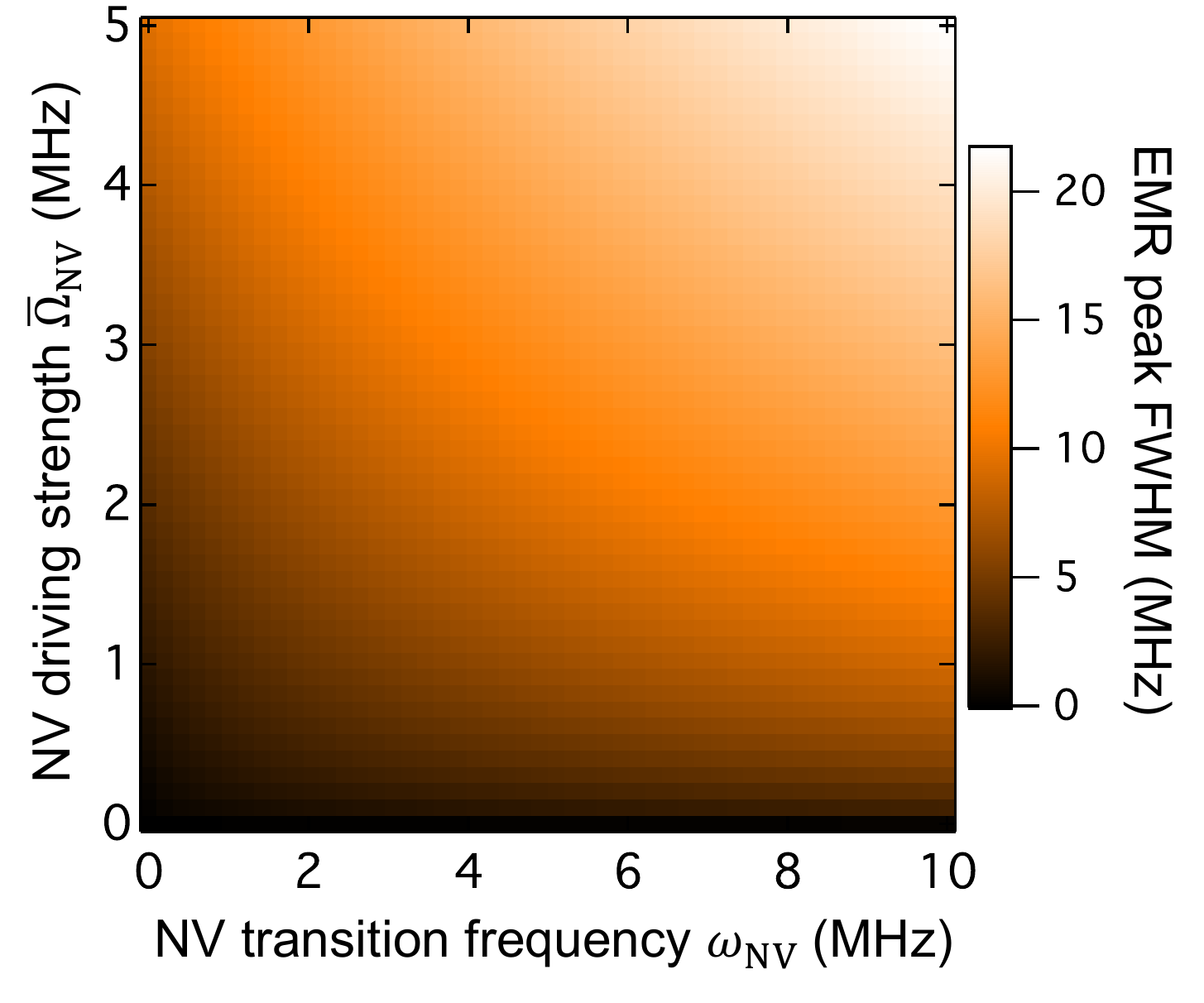}
 		\caption{Numerically solved FWHM of the EMR resonance feature as a function of the NV transition frequency, $\omega_{\nv}$, and the NV driving strength of the NV by the environmental spin ensemble, $\overline{\Omega}_{\nv}$, which gives the EMR driven Rabi frequency of the NV when the environmental spins are driven resonantly, $\omega_{\mw}=\omega_{\es}$, and the EMR matching condition is satisfied, $\Omega_{\es}=\omega_{\nv}$.}
 		\label{FigS3}
 	\end{center}
 \end{figure}
 
 The results, shown in Fig. \ref{FigS3}, are commensurate with the data presented in Figs. $2$ and $3$, giving a FWHM of approximately $2.5$ MHz, for a NV transition frequency, $\omega_{\nv}=2.65$ MHz, and an NV driving strength, $\overline{\Omega}_{\nv}=0.33$ MHz, as extracted from the optimally driven EMR Rabi curve [Fig. $4$a]. The line widths of the resonance features presented in the P1 centre spectrum acquired by EMR spectroscopy [Fig. $5$b] are also commensurate with this analysis, however, the comparatively narrow $200$ kHz line width observed in free-electron resonance peak of Fig. $5$c is problematic. The numerical analysis suggests that such a line width is achievable with $\omega_{\nv}<200$ kHz, and a weak coupling to the environmental spin ensemble, $\overline{\Omega}_{\nv}< 200$ kHz, however this case is not supported by EMR Rabi data from the same NV, which suggests $\overline{\Omega}_{\nv}\approx250$ kHz. This may indicate indicate a breaking of the semi-classical interaction regime assumed by the model and is hence outside of the scope of this analysis.

\end{document}